\date{}
\documentclass[aps,pra,showpacs]{revtex4}
\usepackage{amsfonts}
\usepackage{amsmath}
\usepackage{amssymb}
\usepackage{latexsym}
\usepackage{psfrag}
\usepackage[pdftex]{graphicx} 
\usepackage[center]{caption2}

\begin{document}

\title{Boundary divergences in vacuum self-energies and quantum field theory in curved spacetime}

\author{Francisco D. Mazzitelli$^{1,2}$ \footnote{fdmazzi@cab.cnea.gov.ar}}
\author{Jean Paul Nery$^3$\footnote{jnery@ic.sunysb.edu}}
\author{Alejandro Satz$^4$ \footnote{alesatz@umd.edu}}

\affiliation{$^1$ Centro At\'omico Bariloche,
Comisi\'on Nacional de Energ\'\i a At\'omica,
R8402AGP Bariloche, Argentina}
\affiliation{$^2$ Departamento de F\'\i sica {\it Juan Jos\'e
 Giambiagi}, FCEyN UBA, Facultad de Ciencias Exactas y Naturales,
 Ciudad Universitaria, Pabell\' on I, 1428 Buenos Aires, Argentina}
\affiliation{$^3$ 
Department of Physics and Astronomy,
Stony Brook University, Stony Brook NY 11794  USA
}
\affiliation{$^4$ Maryland Center for Fundamental Physics,
Department of Physics, University of Maryland 
College Park, MD 20742-4111, USA}

\date{today}

\begin{abstract}
It is well known that boundary conditions on quantum fields produce divergences in the 
renormalized energy-momentum tensor near the boundaries. Although irrelevant for the computation
of Casimir forces between different bodies, the self-energy couples to gravity, and
the divergences may, in principle, generate large gravitational effects.
We present an  analysis of the problem in the context of quantum field theory in curved spaces. 
Our model consists of a quantum scalar field coupled to a classical field that, in a certain limit,
imposes Dirichlet boundary conditions on the quantum field. We show that the model is renormalizable and
that the  divergences in the renormalized energy-momentum tensor disappear  for
sufficiently smooth interfaces.
\end{abstract}


\maketitle

\section{Introduction}

The vacuum energy produces measurable forces between neutral bodies \cite{reviews}. But even in the presence of a single body, the modes of the electromagnetic field are disturbed and produce a self-energy. Although irrelevant for the calculation of forces between different objects, this self-energy is in principle observable through its coupling to gravity. 

In the case of perfect conductors, the vacuum energy density, or more generally the energy-momentum tensor, diverges near the boundaries, as noted 
for the first time by Deutsch and Candelas a long time ago \cite{DC}.  These divergences are not the usual ones in quantum field theory, 
because they are present in the {\it already renormalized} energy-momentum tensor. The origin of the divergences is the unphysical
assumption of perfect conductivity for all modes of the electromagnetic field since,  on physical grounds,
one expects any material to become transparent at high energies. More generally,  divergences in the renormalized energy-momentum-tensor are present 
even for non-perfect conductors, as long as there is a sharp boundary between two media with different
electromagnetic properties. In this case the origin of the divergences is that, for 
modes with extremely small wavelengths,  it is unphysical to assume a sharp boundary,
since the transition region between media becomes larger than the wavelength of the high 
frequency modes.  

The problem of the divergences in the self-energies has been considered in flat spacetime in a series of works by Graham et al
\cite{MIT}. It was pointed out  that, if the boundary conditions are replaced by an interaction with a second field, the model is renormalizable and the 
self-energies are not divergent. 
In those works, the authors considered a toy model consisting of a 
vacuum scalar field $\phi$ coupled to
a background (classical field) $\sigma$, with an interaction of the form $\phi^2\sigma$. The sharp 
limit corresponds to consider a discontinuous $\sigma$, and the ``perfect conductor" limit to take 
$\sigma\to\infty$ at a particular point, imposing Dirichlet boundary conditions on the vacuum field at that 
point.  When these limits are taken, the self-energy depends on the ultraviolet cutoff, i.e. on the 
specific interaction between the high energy modes of the quantum field and the microscopic degrees of
freedom of the body.  Recently, Milton \cite{Milton} and Bouas et al \cite{Bouas} considered a similar problem, computing the energy 
density for a scalar vacuum field in some particular potentials (``soft walls"), showing that the energy density
near a potential barrier is finite for sufficiently smooth potentials. For a discussion of some aspects of the coupling of
the Casimir energy to gravity see Ref.\cite{revmilton}.

A complete analysis of the Casimir self-energies and their eventual gravitational implications must be
performed in the context of quantum field theory in curved spacetimes.  
Following Ref.\cite{MIT}, one could
replace the boundary conditions by interactions with a classical field, and show not only that the matter sector of the theory is renormalizable, but also that the usual divergences in the energy-momentum tensor  can be absorbed in the coupling constants of the  gravitational sector, resulting in 
finite and well defined semiclassical Einstein equations when the classical field is sufficiently
smooth. The divergences associated to  the unphysical limits will reappear when considering
discontinuous classical fields. The aim of the present paper is to provide such
analysis. We will consider a scalar vacuum field in curved spacetimes, coupled to a classical field $\sigma$ that models the ``mirror". 
We will show that, for smooth $\sigma$, the model is renormalizable using the standard renormalization procedure for quantum fields in curved spacetimes. 
We will provide some examples to illustrate the appearance  of divergences for sharp interfaces. 

It is worth remarking that similar divergences in the renormalized energy-momentum tensor do appear even for free fields  in curved 
spacetimes, if one assumes that the spacetime metric is not sufficiently smooth. A well known example, in cosmology,  is the 
divergence in the energy density that appears when one considers an abrupt transition between two different epochs (e.g.
De Sitter space to radiation dominated universe \cite{Ford}), that involves a discontinuity in the curvature tensor.
 A less known
example is the divergence of the vacuum polarization around spherically symmetric objects,
at a sharp interface between the object and the vacuum \cite{stars}.

The paper is organized as follows. In the next section we introduce the model. In Section 3 we prove that
the divergences in the energy-momentum tensor can be absorbed into the bare constants of the theory, yielding finite semiclassical Einstein equations. In Section 4 we present some explicit calculations for weakly coupled mirrors, showing that as long as the potential that models the mirror is smooth enough,  there are no divergences in the renormalized energy-momentum tensor and in the renormalized coincidence limit of the two point function, $\langle\phi^2\rangle$.  On the other hand,
discontinuities in the potential or in its first two derivatives produce infinite answers. Section 5 contains a discussion of the main results. We will work with natural units $\hbar =c=1$ and metric signature $+---$.

\section{The Model}

We consider a quantum vacuum field $\phi$ interacting with a background classical field $\sigma$ on a curved spacetime.
The action of the complete system is
\begin{equation}
S=S_{mat}+S_{grav}
\label{eq: accion2}
\end{equation}
where
\begin{equation}
\begin{split}
S_{mat} =  \frac{1}{2}\int d^4x \sqrt{g}\Big( & \phi_{;\mu} \phi^{;\mu}-(m_1^2+\xi_1 R +\tfrac{\lambda_1}{2}\sigma^2) \phi^2 + \\
& + \sigma_{;\mu}\sigma^{;\mu}- (m_2^2+\xi_2 R )\sigma^2 - \frac{\lambda_2}{12} \sigma^4 \Big)
\end{split}
\label{eq: accion de materia}
\end{equation}
and
\begin{equation}
S_{grav} =\frac{1}{2} \int d^4 x \sqrt{g} \left[\frac{1}{\kappa}(R - 2 \Lambda) - \epsilon_1R^2 - \epsilon_2 R_{\mu \nu}R^{\mu \nu} - \epsilon_3 R_{\mu \nu \rho \sigma} R^{\mu \nu \rho \sigma}\right], .
\label{eq: accion gravitatoria}
\end{equation}
Here $R_{\mu \nu}={R^{\alpha}}_{\mu \alpha \nu}$, $\kappa = 8 \pi G $, $\Lambda$ is the cosmological 
constant, and  $\epsilon_i$, i=1,2,3 are dimensionless parameters. The terms quadratic in the curvature are
needed in order to renormalize the theory,  as is the
self-interaction term for $\sigma$. Note that the classical field $\sigma$ provides a position-dependent
mass for the field $\phi$, and therefore the propagation of $\phi$ will be suppressed in regions where the mass
is very high. In this sense $\sigma$ models a ``mirror". For example, a thin mirror located at
$x=x_0$ is described by the interaction
\begin{equation}
\lambda_1\sigma^2\phi^2=\lambda_1\delta(x-x_0)\phi^2\, ,
\end{equation}
and the perfect conductor limit corresponds to $\lambda_1\to\infty$.

The classical field equations are

\begin{equation}
(\Box + m_1^2 + \xi_1 R + \frac{\lambda_1}{2}\sigma^2) \phi = 0
\label{eq: phi}
\end{equation}

\begin{equation}
(\Box + m_2^2 + \xi_2 R + \frac{\lambda_1}{2}\phi^2) \sigma + \frac{\lambda_2}{6}\sigma^3 = 0
\label{eq: sigma}
\end{equation}

\begin{equation}
\frac{1}{\kappa}\tilde{G}_{\mu \nu} = - T_{\mu \nu}^{(\sigma)} - T_{\mu \nu}^{(\phi)}
\label{eq: einstein}
\end{equation}

where

\begin{equation}
\begin{split}
\frac{2}{\sqrt{g}}\frac{\delta S_{grav}}{\delta g^{\mu \nu}} & =  \frac{1}{\kappa}\tilde{G}_{\mu \nu} = \\
& = \frac{1}{\kappa}(R_{\mu \nu} - \frac{1}{2} R g_{\mu \nu} + \Lambda g_{\mu \nu}) + \epsilon_1 H_{\mu \nu}^{(1)} + \epsilon_2 H_{\mu \nu}^{(2)} + \epsilon_3 H_{\mu \nu}
\end{split}
\end{equation}

and 

\begin{equation}
\frac{2}{\sqrt{g}}\frac{\delta S_{mat}}{\delta g^{\mu \nu}}= T_{\mu \nu}^{(\sigma)} + T_{\mu \nu}^{(\phi)}.
\end{equation}
The tensors $H_{\mu\nu}^{(i)}$ come from the variation of the terms quadratic in the curvature contained in the gravitational action.
The classical energy-momentum tensor for the $\sigma$ field is

\begin{equation}
\begin{split}
T_{\mu \nu}^{(\sigma)} & = ( 1 - 2 \xi_2) \sigma_{;\mu} \sigma_{;\nu} + (2\xi_2-\tfrac{1}{2}) g_{\mu \nu} \sigma_{;\rho}\sigma^{;\rho} - 2 \xi_2 \sigma \sigma_{;\mu \nu} +\\
& + 2 \xi_2 g_{\mu \nu} \sigma \Box \sigma - \left[\xi_2 (R_{\mu \nu} - \frac{1}{2} R g_{\mu \nu}) - \frac{g_{\mu \nu}}{2}(m_2^2 + \frac{\lambda_2}{12} \sigma^2)\right]\sigma^2\, ,
\end{split}
\label{eq: Tuv2}
\end{equation}

\noindent while  $ T_{\mu \nu}^{(\phi)} $ is the energy-momentum tensor for a free field with variable mass $m_1^2 + \frac{\lambda_1}{2} \sigma^2$, that is,
\begin{equation}
\begin{split}
T_{\mu \nu}^{(\phi)} & = ( 1 - 2 \xi_1) \phi_{;\mu} \phi_{;\nu} + (2\xi_1-\tfrac{1}{2}) g_{\mu \nu} \phi_{;\rho}\phi^{;\rho} - 2 \xi_1 \phi \phi_{;\mu \nu} +\\
& + 2 \xi_1 g_{\mu \nu} \phi \Box \phi - \left[\xi_1 (R_{\mu \nu} - \frac{1}{2} R g_{\mu \nu}) - \frac{g_{\mu \nu}}{2}(m_1^2 + \frac{\lambda_1}{2} \sigma^2)\right]\phi^2\, .
\end{split}
\label{eq: Tuvphi}
\end{equation}

We consider now the semiclassical version of the theory, in which the field $\phi$ becomes a 
quantum field while $\sigma$ and $g_{\mu\nu}$ are treated classically. The Heisenberg equation
for the quantum operator associated to $\phi$ is given by  the classical Eq.(\ref{eq: phi}). The evolution equations
for the classical backgrounds are obtained by taking the mean value of the classical Eqs.(\ref{eq: sigma}) and
(\ref{eq: einstein}):

\begin{equation}
\frac{1}{\kappa}\tilde{G}_{\mu \nu} = - T_{\mu \nu}^{(\sigma)} - \langle T_{\mu \nu}^{(\phi)}\rangle 
\label{eq: einstein2}
\end{equation}

\begin{equation}
(\Box + m_2^2 + \xi_2 R + \frac{\lambda_1}{2}\langle \phi^2\rangle ) \sigma + \frac{\lambda_2}{6}\sigma^3 = 0
\label{eq: sigma2}
\end{equation}

We have therefore established a well defined model to study the divergences of the self-energy in Casimir calculations. The quantities  $\langle \phi^2\rangle $  and $\langle T_{\mu \nu}^{(\phi)}\rangle $ are formally divergent. The divergences must be absorbed into the bare constants of the theory. The renormalized version of  $\langle T_{\mu \nu}^{(\phi)}\rangle $ contains information about the Casimir effect
(force between different objects), as well as the self-energy. The gravitational effects produced by the vacuum energy
can in principle be computed by considering it as a source 
of the semiclassical Einstein equations.

\section{Renormalizability}

The theory of  interacting fields in curved spacetimes can be renormalized using a precise covariant procedure \cite{books qftcs}. In the present model it is necessary to introduce 
minor modifications to take into account that not only the metric but also one of the interacting fields is treated
classically. 

It will be particularly useful to adapt the renormalization method described
in Ref.\cite{paz}. It is shown there that to analyze the renormalizability of $\lambda\varphi^4$ theory in curved spaces at the level of 
the equations of motion, one can split the field as $\varphi=\varphi_0+\hat\varphi$, where $\varphi_0$ is the mean value of the field
and $\hat\varphi$ is the quantum operator that describes the fluctuations around the mean value. To one loop order,
$\hat\varphi$ satisfies a free field equation with a variable mass. The situation in our model is very similar, since the quantum field
$\phi$ can be thought of a free field with variable mass. Therefore, we can follow closely Ref.\cite{paz}.
 
We define the renormalized quantities
\begin{equation}
\begin{split}
\langle \phi^2\rangle _{ren}  & =  \langle \phi^2\rangle  - \langle \phi^2\rangle _{ad2}\\
\langle T_{\mu \nu}^{(\phi)}\rangle _{ren}  = & \langle T_{\mu \nu}^{(\phi)}\rangle  - \langle T_{\mu \nu}^{(\phi)}\rangle _{ad4}
\end{split}
\label{eq: renormalizacion}
\end{equation}
where $\langle T_{\mu \nu}^{ (\phi)}\rangle _{ad4}$ and $\langle \phi^2\rangle _{ad2}$
are constructed using the Schwinger DeWitt expansion up to fourth and second adiabatic order respectively \cite{adiab}. The divergences
present in these quantities are to be absorbed into the bare constants of the theory.

The usual Schwinger DeWitt expansion for the propagator of a massive field reads \cite{SDW}
\begin{equation}
\begin{split}
 G^{(1)}_{SD}(x,x') & = -2 \mbox{Im} G_F^{SD}(x,x') = \\
& = -2 \mbox{Im} \left[\Delta^{1/2}(x,x')  \int\limits_0^{\infty} \frac{ds}{(4 \pi i s)^{n/2}} e^{\frac{i \sigma(x,x')}{2s}-im^2s} \sum_{j \geq 1}(is)^j \Omega_j(x,x')\right]
\end{split}
\label{eq: SDW2}
\end{equation}
where $G_F$ is the Feynman propagator,  $\Delta(x,x')$ is the  van-Vleck determinant, $\sigma(x,x')$
is one half of the geodesic distance between $x$ and $x'$ and $n$ is the number of spacetime dimensions,
that acts as a regulator. The functions  $\Omega_j(x,x')$ are defined by a set of recursive equations that follows from imposing
the equation for the propagator. 

When the field has a variable mass, this expansion can  be  generalized to \cite{paz}
\begin{equation}
\begin{split}
 G^{(1)}_{SD}(x,x') & = -2 \, \mbox{Im} G_F(x,x') = \\
& = -2 \, \mbox{Im} \left[ \Delta^{1/2}(x,x')  \int\limits_0^{\infty} \frac{ds}{(4 \pi i s)^{n/2}} e^{\frac{i \sigma(x,x')}{2s}-i f(x,x') s} \sum_{j \geq 1}(is)^j \Omega_j(x,x')\right]\, ,
\end{split}
\label{eq: SDW}
\end{equation}
where  $f(x,x') = \frac{1}{2}[M^2(x) + M^2(x')]$ with $M^2(x) = m_1^2 + \frac{\lambda_1}{2} \sigma^2$. 

The quantities to be subtracted to cancel the divergences of  $\langle \phi^2\rangle$ and $\langle T_{\mu \nu}\rangle$
are \cite{paz}
\begin{equation}\label{phi2ImGf}
\langle \phi^2\rangle _{ad2} = - \mbox{Im} \left[ {G_F^{SD}}|_{ad2}\right]
\end{equation}
and
\begin{equation}
\begin{split}
\langle T_{\mu \nu}^{(\phi)}\rangle _{ad4}  = & - 2 \, \mbox{Im}\left\{ -\tfrac{1}{2} ([G_F^{SD}]_{; \mu \nu})_{ad4} + \tfrac{1}{2} (\tfrac{1}{2} - \xi_1) ([G_F^{SD}]_{;\mu \nu})_{ad4} + \right.\\
& \left. - \tfrac{1}{2} (\xi_1- \tfrac{1}{4})g_{\mu \nu} (\Box [G_F^{SD}])_{ad4} - \tfrac{1}{2}\xi_1 R_{\mu \nu}[G_F^{SD}]_{ad2} \right\}\, .
\end{split}
\label{eq: Tuv y G ad}
\end{equation}
We emphasize that we are not using a point-splitting regularization but a dimensional regularization. We will substitute
$\langle T_{\mu \nu}^{(\phi)}\rangle $ and $\langle \phi^2\rangle $ in the semiclassical equations
by $\langle T_{\mu \nu}^{(\phi)}\rangle _{ren}+\langle T_{\mu \nu}^{(\phi)}\rangle _{ad4}$ and $\langle \phi^2\rangle _{ren}+\langle \phi^2\rangle _{ad2}$,  respectively, and absorb the infinities contained in
$\langle T_{\mu \nu}^{(\phi)}\rangle _{ad4}$ and $\langle \phi^2\rangle _{ad2}$
 into the bare constants appearing in those equations.

\subsection{Renormalization of the equation for the classical field}

The coincidence limit of the two point function for a field with variable mass is given by \cite{paz}
\begin{equation}
\langle \phi^2\rangle _{ad2} = \frac{1}{(4 \pi)^2} \left( \frac{M^2}{\tfrac{n}{2}-1} + (\xi_1 - \tfrac{1}{6})R \right) \left(\frac{2}{n-4} + \mbox{ln} \frac {M^2}{\mu^2} + O (n-4) \right)\, ,
\label{eq: phi2 ad2}
\end{equation}
where $\mu$ is an arbitrary constant with dimensions of mass. 
Inserting this expression into the semiclassical equation for $\sigma$ and writing the bare constants
in terms of renormalized ones
\begin{equation}
\begin{split}
m_2^2 & =m_{2R}^2+\delta m_2^2\\
 \xi_2 & = \xi_{2R} + \delta \xi_2\\
\lambda_2 & = \lambda_{2R}+\delta \lambda_2
\end{split}
\label{eq: reescritura}
\end{equation}
we obtain
\begin{equation}
\begin{split}
\Box & \sigma + \Big[m_{2R}^2 + \delta m_2^2 + (\xi_{2R} + \delta \xi_2) R\Big] \sigma + \frac{1}{3!} (\lambda_{2R} + \delta \lambda_2) \sigma^3 + \frac{\lambda_{2R} \,\sigma}{32 \pi^2}\Big[M^2 + \\
\\
& +(\xi_1 - \frac{1}{6}) R \Big] \mbox{ln} \frac{M^2}{\mu ^2} + \frac{\lambda_1 \sigma}{16 \pi^2 (n-4)} \left(\frac{M^2}{\frac{n}{2}-1} + (\xi_1 - \frac{1}{6}) R \right) + \frac{\lambda_1}{2} \sigma \langle \phi^2\rangle _{ren} = 0\, .
\end{split}
\label{eq: sigma3}
\end{equation}
Therefore, the divergences can be absorbed with  the counterterms
\begin{equation}
\begin{split}
\delta m_2^2 & = -\frac{\lambda_1 m_1^2}{16 \pi^2 (\frac{n}{2}-1)(n-4)}+ \Delta m_1^2 \\
\delta \xi_2 & = - \frac{\lambda_1 (\xi_1 - \frac{1}{6})}{16 \pi^2 (n-4)} + \Delta \xi_2 \\
\delta \lambda_2 & =  - \frac{3 \lambda_1^2}{16 \pi^2 \frac{n}{2}(\frac{n}{2}-1)(n-4)} + \Delta \lambda_2
\, ,
\label{eq: redefiniciones1}
\end{split}
\end{equation}
where $\Delta m_2^2, \, \Delta \xi_2$ and $\Delta \lambda_2$ are finite contributions (they vanish in the minimal 
prescription scheme). Note that in the conformal case ($m_1=0$ and $\xi_1=1/6$),  only 
a counterterm for the self-coupling of $\sigma$ is needed.

\subsection{Renormalization of the semiclassical Einstein equation }

Let us now consider the renormalization of the gravitational sector of the theory.  As $n\to 4$, the divergent
part of $\langle T_{\mu \nu}^{(\phi)}\rangle _{ad4}$ is given by \cite{pazappendix}
 \begin{equation}
\begin{split}
\langle T_{\mu \nu}^{(\phi)}\rangle _{ad4}^{div} = & + \frac{1}{8 \pi^2 (n-4)}\Bigg[\frac{m_1^4 g_{\mu \nu}}{n \left (\frac{n}{2}-1\right)} - \frac{m_1^2}{\frac{n}{2}-1}(\xi_1 - \tfrac{1}{6}) G_{\mu \nu} + \tfrac{1}{2}(\xi_1-\tfrac{1}{6})^2 H_{\mu \nu}^{(1)}+ \\ 
& + \left.\tfrac{1}{180}(H_{\mu \nu}-H_{\mu \nu}^{(2)})\Bigg]-\frac{\lambda_1}{16 \pi^2 (n-4)} \bigg[(\xi_1 - \tfrac{1}{6})(\sigma^2_{;\mu \nu} + g_{\mu \nu}\Box \sigma^2) - \right. \\
& - \left. \sigma^2 \left( - \frac{\xi_1 - \frac{1}{6}}{\frac{n}{2}-1} G_{\mu \nu} + \frac{g_{\mu \nu}}{\frac{n}{2}(\frac{n}{2}-1)} (m_1^2 + \tfrac{\lambda_1}{4}\sigma^2)\right)\right]\, .
\end{split}
\label{eq: Tuv div}
\end{equation}
This expression contains both geometric divergences and divergences dependent on  the classical field $\sigma$ (and
its derivatives). The former should be absorbed into a redefinition of the gravitational constants appearing in the
left hand side of the semiclassical Einstein equations, and the latter into a redefinition of the constants associated to the $\sigma$
field, which appear in $T_{\mu \nu}^{(\sigma)} $. 

Inserting Eq.(\ref{eq: Tuv div}) into Eq.(\ref{eq: einstein2}) one can show that the divergences dependent on $\sigma$ can be absorbed using the same counterterms given in Eq.(\ref{eq: redefiniciones1}). This is a non trivial check of our calculations, and a 
necessary condition for the renormalizability of the theory. On the other hand, the geometric divergences
can be absorbed into the gravitational constants, by choosing the following counterterns
\begin{subequations}
\begin{equation}
\epsilon_1 = \epsilon_{1R} - \frac{(\xi_1 - \frac{1}{6})^2}{16 \pi^2 (n-4)} + \Delta \epsilon_1\\
\end{equation}
\begin{equation}
\epsilon_2 = \epsilon_{2R} + \frac{1}{1440 \pi^2 (n-4)} + \Delta \epsilon_2\\
\end{equation}
\begin{equation}
\epsilon_3 = \epsilon_{2R} + \frac{1}{1440 \pi^2 (n-4)} + \Delta \epsilon_3
\end{equation}
\label{eq: redefiniciones2}
\end{subequations}
\begin{subequations}
\begin{equation}
\kappa^{-1}= \kappa_R^{-1} + \frac{1}{8 \pi^2} \frac{(\xi_1 - \tfrac{1}{6})m_1^2}{n-4} + \Delta \kappa^{-1}
\label{eq: kappa1}
\end{equation}
\begin{equation}
\Lambda \kappa^{-1} = (\Lambda \kappa^{-1})_R - \frac{1}{8 \pi^2} \frac{m_1^4}{n (n/2 - 1) (n-4)} + \Delta (\Lambda \kappa^{-1})\, .
\label{eq: kappa2}
\end{equation}
\label{eq: redefiniciones3}
\end{subequations}

This completes the proof of the renormalizability of the model,  and is one of the main results of this paper.
We have shown that, if the presence of a mirror is modeled by the interaction of the vacuum field with a classical 
background field, the divergences in the vacuum expectation value $\langle T_{\mu \nu}^{(\phi)}\rangle$ can be 
absorbed into the bare constants appearing in the semiclassical Einstein equations.  Not only
the gravitational constants are renormalized but also the bare constants associated to the Lagrangian of the classical
background field.  The renormalizability is valid for one or
more mirrors of arbitrary shape, as long as they can be described by a smooth function $\sigma$. 
Note that the divergences in the energy-momentum tensor are considerably simpler in the conformal case $m_1=0$ and
$\xi_1=1/6$.

As a final remark, we stress that the quantities $\langle\phi^2\rangle_{ren}$ and $\langle T_{\mu \nu}^{(\phi)}\rangle_{ren}$
will be finite for a sufficiently smooth background field $\sigma$ and spacetime metric $g_{\mu\nu}$.  We will illustrate this in the next 
section.

\section{Explicit evaluations in the weak field approximation}

In this section we present simple expressions for evaluating explicitly $\langle \phi^2(x)\rangle$ and $\langle T_{\mu \nu}^{(\phi)}(x)\rangle$ for a given background potential $\sigma(x)$, within a weak field approximation. For simplicity we will work in Minkowski space, and set $m_1=0$ so that the quantum field is massless. The sense in which we define the weak field approximation is that we require $M^4\ll\square (M^2)$, or in other words $\lambda_1^2\sigma^4(x)\ll\lambda_1\square(\sigma^2(x))$. This can be achieved by having a weak coupling $\lambda_1$, or a weak and/or rapidly varying field $\sigma(x)$. The calculation will stay at the lowest order in $\lambda_1$.

The procedure we follow is similar to the one used in \cite{stars}. It is based in solving the equation for the Feynman Green function,
\begin{equation}\label{Gf}
\left(\square + \frac{\lambda_1}{2}\sigma^2(x)\right)G_F(x,x')=-\delta(x-x'),
\end{equation}
to first order in $\lambda_1$, and using the resulting $G_F^{(1)}(x,x')$ to compute $\langle \phi^2(x)\rangle^{(1)}$ and $\langle T_{\mu \nu}^{(\phi)}(x)\rangle^{(1)}$ in the coincidence limit. Expanding Eq.(\ref{Gf}) by writing $G_F = G_F^{(0)}+G_F^{(1)}$ in a $\lambda_1$ expansion and discarding the second-order term, we obtain
\begin{equation}
\square\,G_F^{(1)}(x,x')=-\frac{\lambda_1}{2}\sigma^2(x)\,G_F^{(0)}(x,x')\,.
\end{equation}
This is solved explicitly using a momentum-space representation for $G_F^{(0)}$:
\begin{equation}\label{GFpspace}
G_F^{(1)}(x,x')=\frac{\lambda_1}{2(2\pi)^8}\int \mathrm{d}^4 \tilde{x}\int\mathrm{d}^4 k \int \mathrm{d}^4 k'\,\frac{\mathrm{e}^{-i k(x-\tilde{x})}}{k^2}\,\frac{\mathrm{e}^{-i k'(\tilde{x}-x')}}{k'^2}\,.
\end{equation}
After switching variables to $p=k-k'$ and $q=k+k'$, the $k$-integrals can be done by passing to $n$-dimensional space and employing standard dimensional regularization techniques, with the result in the coincidence limit being:
\begin{equation}
G_F^{(1)}(x,x)=-\frac{i}{16\pi^2}\frac{\lambda_1}{n-4}\sigma^2(x)-\frac{i\,\lambda_1}{512\pi^6}\int\mathrm{d}^4 p\int\mathrm{d}^4 \tilde{x}\,\mathrm{e}^{ip(x-\tilde{x})}\sigma^2(\tilde{x})\ln\left(-\frac{p^2}{\mu^2}\right)\,.
\end{equation}
Here $\mu$ is an arbitrary mass scale introduced in the regularization procedure. The divergence in the first term, which is purely local, can be absorbed in the coefficient $\xi_2$ of the action, as can be seen by comparison with the second line of Eq.(\ref{eq: redefiniciones1}). Hence in accordance with Eq.(\ref{phi2ImGf}) we have
\begin{equation}\label{fourierdoble4d}
\langle  \phi^2(x)\rangle_{\mathrm{ren}} = \frac{\lambda_1}{512\pi^6}\int \mathrm{d}^4 p\int\mathrm{d}^4 \tilde{x}\,\mathrm{e}^{ip(x-\tilde{x})}\sigma^2(\tilde{x})\ln\left(-\frac{p^2}{\mu^2}\right)\,.
\end{equation}
For example, in the particular case in which the background potential depends only on the spatial coordinate $z$, we shall have:
\begin{equation}\label{fourierdoble}
\langle \phi^2(z)\rangle_{\mathrm{ren}} = \frac{\lambda_1}{64\pi^3}\int \mathrm{d}p_z \int\mathrm{d} \tilde{z}\,\mathrm{e}^{-ip_z(z-\tilde{z})}\sigma^2(\tilde{z})\ln\left(\frac{p_z^2}{\mu^2}\right)\,.
\end{equation}
This expression can be used to compute $\langle \phi^2(z)\rangle_{\mathrm{ren}}$ for the case of one or several parallel flat mirrors, which are ``almost transparent'' in the sense that the expression is valid to first order in the coupling $\lambda_1$.

We will prove now that according to this model $\langle \phi^2(z)\rangle_{\mathrm{ren}}$ is everywhere finite if the background potential $\sigma^2(z)$ is an integrable $C^2$ function. Under this assumption, it follows as a 
corollary of the Riemann-Lebesgue lemma that the Fourier transform of $\sigma^2(z)$ (which we note by $\hat{\sigma^2}(p_z)$) falls off faster than $p_z^{-2}$. Hence $g(p_z)\equiv\ln|p_z|\hat{\sigma^2}(p_z)$ is integrable, from which it follows that its Fourier transform is in turn well-defined and finite at all points. Since according to (\ref{fourierdoble}) it is precisely the Fourier transform of $g(p_z)$ that gives the nonlocal part of $\langle \phi^2(z)\rangle_{\mathrm{ren}}$, and the local part is the term involving $\ln(\mu^2)\sigma^2(z)$, it follows that both the local and the nonlocal parts of  $\langle \phi^2(z)\rangle_{\mathrm{ren}}$ are finite for all $z$ if $\sigma^2(z)$ is an integrable $C^2$ function.

We have therefore proved that the divergences in $\langle \phi^2\rangle$ are removed if the background field modeling the mirrors is sufficiently well-behaved. We are not able to give a general proof using weaker assumptions than $C^2$ continuity. However, we will see that in simple concrete examples the assumption of $C^0$ continuity is sufficient to obtain finite results.

Let us first consider a discontinuous mirror of width $2z_0$, specified by the background potential:
\begin{equation}\label{discont}
\sigma^2(z)=\frac{1}{2\,L\,z_0}\Theta(z_0-|z|)\,,
\end{equation}
where $\Theta(z)$ is the unit step function. The parameter $L$ has length dimension and is introduced so that $\sigma^2(z)$ is normalized by $\int \sigma^2(z)\, \mathrm{d}z=L^{-1}$. Introducing this expression into Eq.(\ref{fourierdoble}) we obtain as a result:
\begin{equation}\label{phi2dis}
\langle \phi^2(z)\rangle_{\mathrm{ren}} = -\frac{\lambda_1}{64\pi^2 L z_0}\left[\mathrm{sg}(z+z_0)\ln(|z+z_0|+\gamma)-\mathrm{sg}(z-z_0)\ln(|z-z_0|+\gamma)\right]
\end{equation}
where we omit the local contribution coming from the $\ln(\mu^2)$ term. Notice that the result diverges logarithmically at the boundary of the mirror, where $\sigma^2(z)$ is discontinuous.

Consider next a continuous mirror given by
\begin{equation}\label{mirrorcont}
\sigma^2(z) = \frac{1}{L(2z_0+d)}\times \left\{
\begin{array}{rl}
0 & \text{if }\quad z < -(z_0+d),\\
1+\frac{z_0+z}{d} & \text{if } \quad-(z_0+d)<z<-z_0,\\
1 & \text{if }\quad -z_0<z<z_0,\\
1+\frac{z_0-z}{d} & \text{if }\quad z_0<z<z_0+d,\\
0 & \text{if }\quad z > z_0+d,
\end{array} \right.
\end{equation}
wherein the discontinuities in Eq.(\ref{discont}) are resolved with linear interpolations of width $d$. For this mirror the vacuum polarization is computed to be:
\begin{align}\label{phi2cont}
\langle \phi^2(z)\rangle_{\mathrm{ren}} &= \frac{\lambda_1}{32\pi^2 L}\frac{1}{d(2z_0+d)}\left[|z+z_0|\ln|z+z_0|+|z-z_0|\ln|z-z_0|\right.\nonumber\\
&\left.-|z+z_0+d|\ln|z+z_0+d|-|z-z_0-d|\ln|z-z_0-d|\right],
\end{align}
where we have omitted again a local contribution proportional to $\sigma^2$. This result is everywhere finite; it can be compared to that of Eq.(\ref{phi2dis}) in the plot in Figure 1.

\begin{figure}[htp]
\begin{center}
\includegraphics[width=0.8\textwidth]{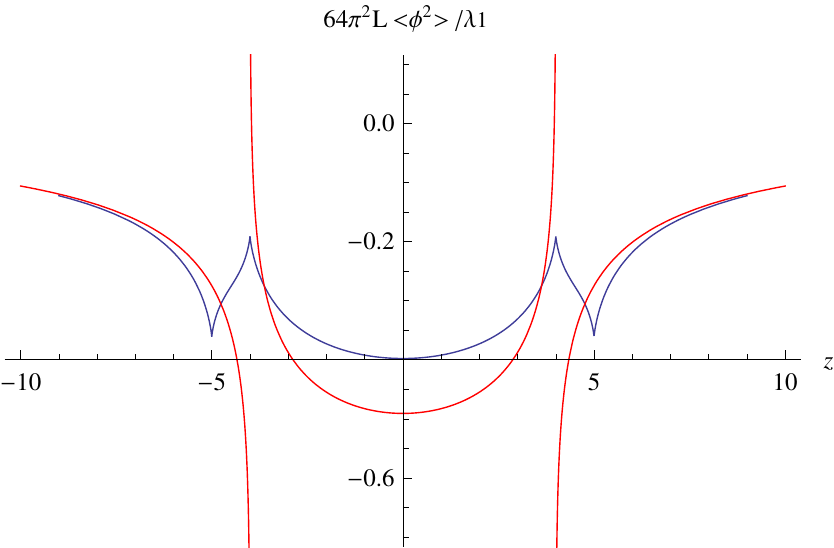}
\end{center}
\caption{A comparison of $\langle \phi^2(z)\rangle_{ren}$ for a discontinuous potential (red) and a continuous one (blue), with $z_0=4$ and $d=1$. Note that  $\langle \phi^2(z)\rangle_{ren}$ diverges at the points where the background field $\sigma$
is discontinuous, and that the divergences disappear when $\sigma$ is continuous.}
\label{fig:phi2plot}
\end{figure}

We turn now to the study of $\langle T_{\mu \nu}^{(\phi)}(x)\rangle$ to first order in $\lambda_1$. From the expression (\ref{eq: Tuvphi}) it follows that  $\langle T_{\mu \nu}\rangle$ (we drop the $\phi$ label from now on) is obtained from the coincidence limit of $G_F(x,x')$ as:
\begin{align}
\langle & T_{\mu \nu}(x)\rangle = -\mathrm{Im}\Bigg[\left(1-2\xi_1\right)\partial_\mu \partial'_\nu G_F(x,x')+\left(2\xi_1-\frac{1}{2}\right)\eta_{\mu\nu}\eta^{\rho\sigma} \partial_\rho \partial'_\sigma G_F(x,x')\nonumber\\
&  - 2\,\xi_1\,\partial_\mu \partial_\nu G_F(x,x')+2\,\xi_1\,\eta_{\mu\nu}\eta^{\rho\sigma} \partial_\rho \partial_\sigma  G_F(x,x') +\frac{\lambda_1}{4}\eta_{\mu\nu}G_F(x,x')\Bigg]\Bigg|_{x=x'}\,.
\end{align}
The coincidence limit of the last term is $\sim \lambda_1 \langle \phi^2\rangle$, and since $\langle \phi^2\rangle$ is $O(\lambda_1)$ we may discard this term in our first-order approximation. The coincidence limit of the remaining term can be performed in an analogous way to that of $\langle \phi^2\rangle$, by passing to $n$-dimensional space, using the momentum space representation (\ref{GFpspace}) for the first-order $G_F(x,x')$, and using standard dimensional regularization techniques. The divergences appearing are proportional to $\sigma^2,_{\mu\nu} + \eta_{\mu\nu}\square \sigma^2$, and can be absorbed into a redefinition of $\xi_2$ as we did before. After the renormalization is performed and we return to $n=4$ we are left with
\begin{align}\label{TmunuRen}
\langle T_{\mu \nu}(x)\rangle_{\mathrm{ren}}&=\frac{\lambda_1}{512\pi^6}\left(\xi_1-\frac{1}{6}\right)\int \mathrm{d}^4p \int\mathrm{d}^4 \tilde{x}\,\mathrm{e}^{ip(x-\tilde{x})}\sigma^2(\tilde{x})\ln\left(-\frac{p^2}{\mu^2}\right)\nonumber\\
&\times \left(p_\mu p_\nu-p^2\eta_{\mu\nu}\right)\,,
\end{align}
which is analogous to (\ref{fourierdoble4d}) for $\langle \phi^2(x)\rangle_{\mathrm{ren}}$. For the case in which $\sigma^2$ depends only on $z$ we get by analogy with (\ref{fourierdoble}):
\begin{align}\label{fourierdobleTmunu}
\langle T_{\mu \nu}(z)\rangle_{\mathrm{ren}}&=\frac{\lambda_1}{64\pi^3}\left(\xi_1-\frac{1}{6}\right)\int \mathrm{d}p_z \int\mathrm{d} \tilde{z}\,\mathrm{e}^{-ip_z(z-\tilde{z})}\sigma^2(\tilde{z})\ln\left(\frac{p_z^2}{\mu^2}\right)\nonumber\\
&\times p_z^2 \left(\delta_{\mu z}\delta_{\nu z}+\eta_{\mu\nu}\right)\,.
\end{align}

Comparing with our arguments showing finiteness for $\langle \phi^2(z)\rangle_{\mathrm{ren}}$ 
it is clear that the components of $\langle T_{\mu \nu}(z)\rangle_{\mathrm{ren}}$ will not diverge as long as $h(p_z)\equiv p_z^2\ln|p_z|\hat{\sigma^2}(p_z)$ is integrable. To ensure this it is sufficient to require $\sigma^2(z)$ to be an integrable $C^4$ function, since then  its Fourier transform falls off faster than $p_z^{-4}$. However, for the simple particular example of a mirror similar to (\ref{mirrorcont}) but with polynomials with higher degree of continuity interpolating between $z_0$ and $z_0+d$, we have checked that $C^2$ continuity of $\sigma^2(z)$ is enough to render the result finite.

We stress at this point that the results obtained in this section have their gravitational counterparts:
as shown in Ref.\cite{stars}, when one computes 
$\langle \phi^2(z)\rangle_{\mathrm{ren}}$ and  $\langle T_{\mu \nu}(z)\rangle_{\mathrm{ren}}$
for a free quantum field in a curved background, both quantities diverge at the points where the background is not sufficiently smooth. The examples described in Ref.\cite{stars} refer to the vacuum polarization around spherically symmetric objects, and divergences show up for sharp interfaces, where the matter density is discontinuous.   The divergences disappear when
the matter density and its first two derivatives are continuous across the interface. Similarly, in a cosmological context,
divergences in the renormalized energy-momentum tensor are removed when one replaces an abrupt transition
of the scale factor by a smoother transition in which the scale factor and its two first derivatives are continuous.
The concrete example of the transition between the inflationary period and radiation domination is discussed 
in detail in Ref. \cite{Ford}.

 High vacuum energy densities in the presence of boundaries may potentially produce  large gravitational effects. In our model, the usual divergences are rendered finite when the sharp boundary is replaced by a sufficiently smooth background field. However, the quantum vacuum energy does attain high values if the background field varies over short distances. 
We can make a crude estimation of the maximum value of the energy density using dimensional analysis. From 
Eq.(\ref{fourierdobleTmunu}), it is not difficult to see that if $\sigma^2(z)$ is varying over distances of order $d$ then  we should expect the
maximum value of
 $\langle T_{\mu \nu}(z)\rangle_{\mathrm{ren}}$ to be of order $\lambda_1 \sigma_0^2 / d^2$, where $\sigma_0^2$ is a representative value attained by $\sigma^2(z)$.
The quantity  $\lambda_1 \sigma_0^2$ has units of (length)$^{-2}$, and determines the properties of the mirror. Therefore it will be  identified with the square of the  plasma frequency $\omega_p$. Restoring $\hbar$ and $c$ factors,
 the mass density associated to the vacuum fluctuations is
 of order $\hbar\omega_p^2/c^3d^2$.
If we take $d$ to be in the range of the $10^{-10}$ meters, representing a smearing of the conductor's sharp boundary over atomic length scales, and $\omega_p$ to be in the range of $10^{15}$ Hz, a typical value of the plasma frequency, then we see that firstly, the first-order approximation assumed in this section is validated, and secondly, the peak values of the quantum energy densities at the boundary of the conductor are of order $10^{-13} \mathrm{g/cm^3}$, too small to be gravitationally detectable.

\section{Conclusions}

In this paper we analyzed the coupling of the vacuum self-energy to gravity. We considered a 
model in which the presence of the bodies that disturb the modes of the quantum
fields is described by the interaction with a background classical field.
We have shown that the divergences in the energy-momentum tensor of the quantum fields 
are consistent with the semiclassical Einstein equations, that is, they can be absorbed into the bare constants of the theory. As expected, the divergences in the renormalized energy-momentum tensor
noticed in previous works only appear when considering unphysical limits of perfect conductivity
and/or sharp interfaces, that in our model would correspond to take a non-smooth 
background field. We have shown that modeling the mirrors in the Casimir effect with a background potential $\sigma^2$ removes the divergences attached to the boundaries, as long as  $\sigma^2$ is a sufficiently smooth integrable function 
($C^2$ continuity is sufficient for $\langle \phi^2\rangle_{\mathrm{ren}}$ and $C^4$ continuity for $\langle T_{\mu \nu}\rangle_{\mathrm{ren}}$, though weaker assumptions are enough in simple examples). The proof is carried to first order in the coupling $\lambda_1$.  These results are in tune with those found in Refs.\cite{Milton,Bouas}, where finiteness of the results is shown with exact calculations for some particular examples of $\sigma^2$,
and are analogous to those in Refs.\cite{Ford,stars} for the vacuum polarization 
of free fields in curved backgrounds, when the metric and its first two derivatives are  continuous. 
We expect that similar conditions for finiteness can be obtained beyond the perturbative approximation.

We have considered a toy model for a vacuum scalar field. We have presented some specific examples to illustrate the 
appearance of divergences 
for sharp interfaces. The examples suggest that the gravitational effects
of the self-energy near mirrors are extremely small for realistic values of the relevant parameters. 
We expect similar results to be valid
for more realistic models involving the electromagnetic field.

\section*{Acknowledgements}
This work was supported by UBA, CONICET and ANPCyT.


\begin{thebibliography}{bib}

\bibitem{reviews}
P. W. Milonni, {it The Quantum Vacuum}, Academic Press, San Diego, 1994;
M. Bordag, U. Mohideen, and V. M. Mostepanenko, Phys. Rep.
\textbf{353}, 1 (2001); K. A. Milton, {\it The Casimir Effect:
Physical Manifestations of the Zero-Point Energy} (World
Scientific, Singapore, 2001); S. Reynaud {\it et al.}, C. R. Acad.
Sci. Paris \textbf{IV-2}, 1287 (2001); K. A. Milton, J. Phys. A:
Math. Gen. \textbf{37}, R209 (2004); S.K. Lamoreaux, Rep. Prog.
Phys. \textbf{68}, 201 (2005);
M. Bordag, G.L. Klimchitskaya, U. Mohideen, and V. M. Mostepanenko, {\it Advances in the Casimir Effect},
Oxford University Press, Oxford, 2009.

\bibitem{DC}
D.~Deutsch, P.~Candelas,
  Phys.\ Rev.\  {\bf D20}, 3063 (1979).

\bibitem{MIT}  
 N.~Graham, R.~L.~Jaffe, V.~Khemani, M.~Quandt, M.~Scandurra, H.~Weigel,
  Nucl.\ Phys.\  {\bf B645}, 49-84 (2002); ibidem
  Phys.\ Lett.\  {\bf B572}, 196-201 (2003); ibidem
  Nucl.\ Phys.\  {\bf B677}, 379-404 (2004).

\bibitem{Milton} K. Milton, arXiv:1107.4589.

\bibitem{Bouas} J.D. Bouas, S.A. Fulling, F.D. Mera, K. Thapa, C.S. Trendafilova, and J. Wagner, arXiv:1106.1162.

\bibitem{revmilton}
K. A. Milton, in Lect. Notes Phys. {\bf 834} (2011). Edited by D. Dalvit, P. Milonni, D. Roberts, and F. da Rosa.

\bibitem{Ford} L.~H.~Ford,
  Phys.\ Rev.\  {\bf D35}, 2955 (1987).
  
\bibitem{stars} 
A.~Satz, F.~D.~Mazzitelli, E.~Alvarez,
  Phys.\ Rev.\  {\bf D71}, 064001 (2005).

\bibitem{books qftcs}
N.~D.~Birrell, P.~C.~W.~Davies,
``Quantum Fields In Curved Space,''
Cambridge University Press (1982);
S.A. Fulling, ``Aspects of Quantum Field Theory in Curved Spacetime", London Mathematical Society Student 
Texts (1989);
L. Parker and D. Toms,  ``Quantum Field Theory in Curved Spacetime: Quantized Fields and Gravity",
Cambridge University Press (2009).


\bibitem{paz}  J.~P.~Paz, F.~D.~Mazzitelli,
  Phys.\ Rev.\  {\bf D37}, 2170-2181 (1988).


\bibitem{adiab} The adiabatic order is usually defined as the number of derivatives of the metric tensor. In this case, and as discussed in Ref.\cite{paz}, the definition is generalized as follows: in terms containing derivatives of the classical field $\sigma$, the adiabatic order is the number of derivatives of $\sigma$ plus the exponent
of $\sigma$.

\bibitem{SDW}  S.~M.~Christensen,
  Phys.\ Rev.\  D {\bf 14}, 2490 (1976).
  
\bibitem{pazappendix} See Ref.\cite{paz}, Appendix B.

\end{thebibliography}
\end{document}